\documentclass[10pt, conference, letterpaper]{IEEEtran}
\IEEEoverridecommandlockouts
\usepackage{algpseudocode}
\usepackage{lipsum}
\usepackage{graphicx,dblfloatfix}
\usepackage[export]{adjustbox}
\usepackage{algorithm}
\usepackage{cite}
\usepackage{amsfonts}
\usepackage{textcomp}
\usepackage{graphicx}
\usepackage{epstopdf}
\usepackage{amsmath}
\usepackage[mathscr]{euscript}
\usepackage{amssymb}
\usepackage{epsfig}
\usepackage{color}
\usepackage[rgb]{xcolor}
\usepackage{amsthm}
\usepackage{hyperref}
\usepackage{float}
\usepackage{cases}
\usepackage{multirow}
\usepackage{commath}
\usepackage{dsfont}
\usepackage{optidef}
\usepackage{mathtools}
\usepackage[bottom]{footmisc}
\usepackage{stackengine}
\usepackage{algorithm, algorithmicx, algpseudocode}
\usepackage{subcaption}
\usepackage{caption}
\usepackage{soul}

\newcommand\StateX{\Statex\hspace{\algorithmicindent}}

\algnewcommand{\Inputs}[1]{%
  \State \textbf{Inputs:}
  \Statex \hspace*{\algorithmicindent}\parbox[t]{.8\linewidth}{\raggedright #1}
}
\algnewcommand{\Initialize}[1]{%
  \State \textbf{Initialize:}
  \Statex \hspace*{\algorithmicindent}\parbox[t]{.8\linewidth}{\raggedright #1}
}

\ifCLASSINFOpdf
\else
\fi
\newtheorem{thm}{Theorem}[section]
\newtheorem{lem}[thm]{Lemma}

\algrenewcommand\algorithmicindent{1.0em}

\newcommand{\simulator}
{\textsf{\small{\mbox{\textit{xeoverse{}~}}}}}

\def\BibTeX{{\rm B\kern-.05em{\sc i\kern-.025em b}\kern-.08em
    T\kern-.1667em\lower.7ex\hbox{E}\kern-.125emX}}
    
\begin{document}

\title{Time-Dependent Network Topology Optimization for LEO Satellite Constellations
}

 \author{\IEEEauthorblockN{Dara Ron}
 \IEEEauthorblockA{
 \textit{NC State University, Raleigh, USA}\\
 dron@ncsu.edu}
 \and
 \IEEEauthorblockN{Faisal Ahmed Yusufzai}
 \IEEEauthorblockA{
 \textit{George Mason University, Fairfax, USA}\\
 fyusufza@gmu.edu}
\and
 \IEEEauthorblockN{Sebastian Kwakye }
 \IEEEauthorblockA{
 \textit{George Mason University, Fairfax, USA}\\
 skwakye@gmu.edu}
 \and
 \IEEEauthorblockN{Satyaki Roy}
 \IEEEauthorblockA{
 \textit{The University of Alabama in Huntsville, USA}\\
 sr0215@uah.edu}
 \and
 \IEEEauthorblockN{Nishanth Sastry}
 \IEEEauthorblockA{
 \textit{University Of Surrey, Guildford, UK}\\
 n.sastry@surrey.ac.uk}
 \and
 \IEEEauthorblockN{Vijay K. Shah}
 \IEEEauthorblockA{
 \textit{NC State University, Raleigh, USA}\\
 vijay.shah@ncsu.edu}
 }

\maketitle

\begin{abstract} 
Today's Low Earth Orbit (LEO) satellite networks, exemplified by SpaceX's Starlink, play a crucial role in delivering global internet access to millions of users. However, managing the dynamic and expansive nature of these networks poses significant challenges in designing optimal satellite topologies over time. In this paper, we introduce the \underline{D}ynamic Time-Expanded Graph (DTEG)-based \underline{O}ptimal \underline{T}opology \underline{D}esign (DoTD) algorithm to tackle these challenges effectively. We first formulate a novel space network topology optimization problem encompassing a multi-objective function -- maximize network capacity, minimize latency, and mitigate link churn -- under key inter-satellite link constraints. Our proposed approach addresses this optimization problem by transforming the objective functions and constraints into a time-dependent scoring function. This empowers each LEO satellite to assess potential connections based on their dynamic performance scores, ensuring robust network performance over time without scalability issues. Additionally, we provide proof of the score function's boundary to prove that it will not approach infinity, thus allowing each satellite to consistently evaluate others over time. For evaluation purposes, we utilize a realistic Mininet-based LEO satellite emulation tool that leverages Starlink's Two-Line Element (TLE) data. Comparative evaluation against two baseline methods -- Greedy and $+$Grid, demonstrates the superior performance of our algorithm in optimizing network efficiency and resilience.

\end{abstract}

\section{Introduction} \label{sec1}

The terrestrial 5G network has proven successful in achieving ultra-low latency of $1$ ms and very high internet speeds, with peak data rates of up to $20$ Gbps (downlink) and $10$ Gbps (uplink), and user experience data rates of $100$ Mbps (downlink) and $50$ Mbps (uplink) \cite{ref1}. Despite these advancements, challenges remain for 5G terrestrial networks. The high deployment cost causes fewer installations of 5G base stations (BSs) in rural areas. Additionally, installing 5G terrestrial networks can be impossible or difficult in harsh environments, such as wilderness areas, oceanic regions, and isolated mountains \cite{ref2}. This demonstrates that the 5G terrestrial network is not yet capable of providing high data rates and low-latency access for users anywhere and anytime.

Non-terrestrial networks (NTN) have emerged as a promising solution to the limitations of terrestrial 5G networks. NTN refers to space network technology, which deploys satellites in low-Earth orbit (LEO) (approximately $160$ km to $2,000$ km above the Earth's surface) to provide network access for users in the air, on the sea, and on the ground around the globe \cite{ref2,raman2023dissecting}. LEO satellites not only offer ubiquitous coverage but also feature low launch costs and simplicity of deployment \cite{ref5}, thus it becomes increasingly attractive to world-class companies launching their LEO satellite constellation projects such as Starlink, OneWeb, Amazon Kuiper, Telesat, Lightspeed, and Hongyan \cite{ref3,ref4,ref6}. The leading space network service provider, Starlink, has already launched over $6,078$ satellites in LEO, of which $6,006$ are operational as of May 2024, with a total of $3$ million subscribers. Starlink plans to deploy up to $42,000$ satellites to construct a LEO Mega-constellation, as reported by Space.com \cite{b11}. OneWeb, the second leading LEO provider, has successfully launched $633$ LEO satellites as of Feb 2024, as reported by Spacenews.com \cite{ref7}. Meanwhile, other companies have launched few or no operational LEO satellites yet, but they have plans to deploy them in the near future. Despite the success of space networks achieved by Starlink, it has also introduced numerous research challenges that require attention~\cite{kassem2022browser}. Each satellite can connect to any other within its range and visibility, prompting recent research to focus on topology designs for LEO satellites that optimizes end-to-end network capacity and reduced latency \cite{b15, b16, b17, b18, b4,b26}. 

\begin{figure}[t!]
    \centering
    \includegraphics[width=8cm]{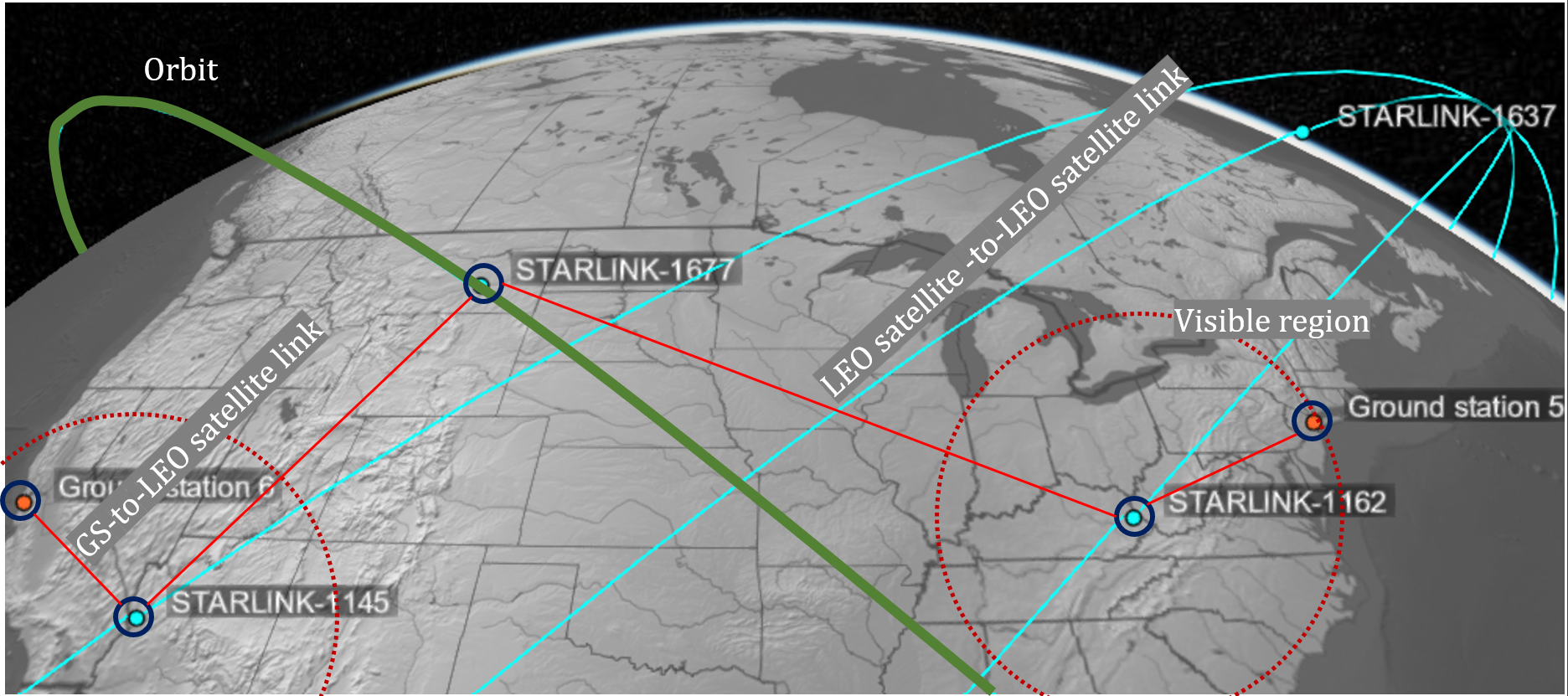}
    \caption{SpaceX's Starlink LEO satellite constellation}
    \label{fig1}
    \vspace{-0.2in}
\end{figure}

Inter-Satellite Links (ISLs) are poised to revolutionize space communication by enhancing data transmission speed and reliability \cite{b12}. The topology designs are based on the ISLs, which are established directly between satellites in orbit, thus allowing them to communicate with each other without relaying signals through GSs. ISLs play a crucial role in modern satellite networks, particularly in constellations and systems designed for global coverage with low latency \cite{b13, b14}. They are currently being used in SpaceX's Starlink satellite network, which comprises thousands of broadband Internet satellites \cite{b25}. The $+$Grid topology is considered the default design for ISLs in Low Earth Orbit (LEO) satellite networks. In this topology, each satellite is connected to four nearby neighbors: two in the same orbital plane and two in adjacent orbital planes \cite{ref8}. This Grid-like structure is relatively easy to design and implement; however, it suffers from significant drawbacks, such as shorter ISL ranges that increase ISL hop counts and end-to-end latency, a lack of consideration for the source and destination nodes when designing the topology, and insufficient flexibility to achieve optimal routing, all of which can affect the overall efficiency of the network. The $\times$Grid has been designed to provide efficiency improvement over the $+$Grid. Similar to the $+$Grid, this approach connects satellites in the same and adjacent orbital planes \cite{b26}. The key difference is that the $+$Grid connection is based on a star topology, whereas the $\times$Grid is based on a rectangular topology. This design minimizes connection overlaps and reduces ISL hop counts by an average factor of two. However, similar to traditional grid topologies, it applies a consistent connection pattern, which limits flexibility in dynamic scenarios. \cite{b4} proposed a topology design based on motifs to improve performance over the neighbor-grid topology. This approach utilizes frequently repeated patterns of connections, where each satellite is connected to its neighbors in the same manner, effectively reducing expensive link changes over time. With this design, performance metrics such as latency and capacity are improved by a factor of two compared to the neighbor grid. However, this approach still relies on the neighbor-based connection strategy, which lacks the flexibility to meet an optimal routing. 

The topology-based neighbor strategies may work perfectly with an ideal satellite constellation to achieve optimal routing paths. For instance, LEO satellites should be deployed in orbital planes with the same or nearly the same altitude, and the inclination and Right Ascension of Ascending Node (RAAN) of the orbits should be equally spaced. This configuration optimizes the distance between parallel orbits and ensures that the distance between satellites within the same orbit is equally spaced, thus achieving optimal routing paths. However, achieving this ideal LEO satellite constellation is challenging. Based on real-time Two-Line Element (TLE) statistical data for LEO satellite experiment, the constellation is not ideal \cite{ref9}. Not all orbital planes are parallel; some orbits intersect with others, resulting in satellites moving in non-parallel directions. The inclinations and RAANs of all orbits are not equally spaced, causing some orbits to be very close to each other, which increases hop counts for grid-based topologies. Additionally, the number of satellites in an orbit and their spacing are not evenly divided, and the perigees of orbits are not the same or nearly the same. These factors indicate that the topology design based on neighbor connections in the same and adjacent orbital planes may not always yield the optimal solution for real satellite constellations. Instead of selecting connections solely between neighbors, the authors in \cite{ref8} conducted an in-depth investigation of LEO satellite topology. They analyzed key parameters that impact network performance, such as the number of orbits, the number of satellites per orbit, and inclination. This approach allows for connections between satellites with the most representative metrics, including maximum round-trip time, geodesic slowdown, and a low number of path changes. The ISL-based dynamic routing algorithm selects the optimal path that maximizes the utility function \cite{ref10}. This algorithm improves key performance metrics (KPMs) such as packet drop rate, end-to-end delay, and throughput. Additionally, \cite{ref11} leveraged deep reinforcement learning (DRL) to interact with satellite networks and select the optimal topology, which minimizes latency across all links. This approach achieves performance improvements in terms of latency, outperforming topology-based neighbor strategies like Motifs and $+$Grid by up to 8.48\% and 42.86\%, respectively. 

\underline{\textit{Research Challenge and Motivation (RCM):}} The challenge remains for neighbor-grid designs due to the imperfect LEO satellite constellations discussed above. The goal of grid-based topologies is to maximize satellite network performance. The research question should be why we select the nearest neighbors rather than focusing on selecting links based on their specific performance metrics. Selecting optimal paths according to their performance metrics can address concerns regarding the ideal constellation, which requires equal spacing between orbital planes, equally spaced LEO satellites within the same orbit, and similar or nearly identical perigees of orbits. This motivation drives the studies in \cite{ref8, ref10, ref11} to develop various methods to select the topology that can directly maximize the KPMs of entire links, such as latency, link churn, packet loss rate, throughput, and utility function. However, a drawback of these previous studies is the lack of consideration for time-dependent performance, which is affected by the dynamic movement of satellites over time. These studies focus solely on selecting satellites based on their performance at each time step. Given the fast movement of LEO satellites, which can reach speeds of up to $7.66$ km/s, the satellite selected at a previous time step may move far away from others or may select the next satellite that does not align with the optimal routing path from source to destination. Therefore, a novel research topic revolves around proposing an optimal topology design strategy that consistently selects satellites with high performance all the time. 

In this study, we propose (i) a \underline{D}ynamic Time-Expanded Graph (DTEG)-based \underline{O}ptimal \underline{T}opology \underline{D}esign (DoTD) algorithm to optimize the time-variant satellite topology 
that maximizes network capacity and simultaneously, minimizes latency and link churn, even with the dynamic changes in space networks due to the fast movement of LEO satellites. Expanding on the previously developed \simulator simulator~\cite{kassem2024xeoverse}, we released SpaceNet simulator as well as the source code and datasets to the community for further research \cite{SpaceNet}. The contributions of our study are outlined as follows.

 $\bullet$ We analyze the visibility of LEO satellite by considering the elevation angle (EA) and key orbital parameters, including the argument of perigee, true anomaly, right ascension of the ascending node (RAAN), and inclination. Based on this analysis, we formulate a new \textit{space network topology optimization} problem, where the multi-objective functions are network capacity, latency, and link churn. The optimization is subject to constraints, such as the visibility of LEO satellites, communication distance, ISL duplex, and the limited number of links that each satellite can select to create the network.

 $\bullet$ We propose a novel DoTD algorithm to address the space network optimization problem. Using DoTD, the optimization problem is transformed into a time-dependent scoring function. This function is derived by normalizing the multiple objective functions to a common scale ranging from $0$ to $1$. This approach enables each LEO satellite to evaluate and choose other satellites for establishing the connections based on their achievable scores. 
 
$\bullet$ We provide proofs demonstrating that: (1) the design of the score function remains finite, thus ensuring its validity for evaluation over time without limitations, (2) the proposed DoTD algorithm consistently achieves optimal performance, and (3) the time complexity of the DoTD algorithm is approximate to $\mathcal{O}(M^2)$.

 $\bullet$ Finally, we augment our proposed DoTD algorithm with a \textit{Open Shortest Path First (OSPF) routing} algorithm to further optimize the routing path for communication between GS sources and destinations. 

\textit{\underline{Experiment}:} We experimentally evaluate by enhancing \simulator~\cite{kassem2024xeoverse} with the proposed DoTD algorithm along with two other baseline methods, Greedy and $+$Grid. Results show that the DoTD achieves an average capacity increase of 28.09\%, a hop count decrement of 10.91\%, and a latency reduction of 39.71\% compared to the Greedy method. Surprisingly, it enhances capacity by up to 70.47\%, reduces hop count by 81.82\%, and decreases latency by up to 96.61\% compared to the $+$Grid method. Additionally, DoTD  significantly minimizes the link churn compared to those of greedy and $+$Grid, significantly improving service continuity and stability. 

\section{Space Network Model}


\subsection{Network Scenario}

This study employs Starlink LEO Constellation \cite{ref9} as the exemplary LEO satellite constellation network, as illustrated in Fig. \ref{fig1}. Multiple satellites are deployed in space at low Earth orbital altitude and move along the orbital planes. Each orbital plane is defined by key six parameters: (1) inclination, (2) RAAN, (3) argument of the perigee, (4) eccentricity, (5) semi-major axis, and (6) true anomaly. Detailed information regarding the key orbital parameters can be found in \cite{ref12}. 

Let $M$ and $N$, where $M \ge N$, denote the number of LEO satellites and orbital planes, respectively. We deploy multiple GSs on Earth, from which two located in different U.S. states or countries are selected to evaluate the satellite network topologies derived from from DoTD (and baselines) topology design algorithms. Let $K$ be the number of GSs. These GSs can communicate with LEO satellites within their visible region. With thousands of LEO satellites launched by SpaceX's Starlink, multiple satellites should be visible to each GS. In this scenario, the source GS selects the satellite with the shortest path to forward its packets to the next satellite and continues this process until the packets reach the GS destination. Due to the fast movement of LEO satellites relative to Earth's gravity, the selected satellite may move out of the visible region. Consequently, the GSs must promptly switch their connections to other satellites. The space networks should consist of two types of communication links: GS-to-LEO satellite and ISLs. 

\subsection{Network Model for GS-to-LEO Satellites} \label{subIIB}

Communication between the GS and the LEO satellite can only be established within the visible region of satellite. This visible region is determined by the minimum and maximum EA, denoted as $\theta_{\min}$ and $\theta_{\max}$, respectively. According to \cite{ref13}, the mathematical relationship between the minimum and maximum EA is expressed as follows:
\begin{equation}
    \theta_{\max} = \arccos\left( R_e \cos (\theta_{\min})/H_{n,t} \right) - \theta_{\min},
\end{equation}
where $R_e$ represents the Earth's radius ($R_e = 6378$ km), and $H_{n,t}$ is the altitude from the Geocenter to the $n$-th orbit at time step $t$, $\forall n = \{1,\dots,N\}$. Given this visibility constraint, we introduce the indicator function $\mathbb{I}_{k,t}$ to represent the possible communication between the $k$-th GS and all LEO satellites. It is defined as $\mathbb{I}_{k,t} = \{\mathbf{1}_{\{\theta_{k,m,t} \in [\theta_{\min}, \theta_{\max}]\}} | m = 1, \dots,M\}$, where $\theta_{k,m,t}$ represents the EA between the $k$-th GS and the $n$-th LEO satellite. The indicator function $\mathbf{1}_{\{\theta_{k,m,t} \in [\theta_{\min}, \theta_{\max}]\}}$ equals to $1$ when the $k$-th GS is within the visible region $(\theta_{k,m,t} \in [\theta_{\min}, \theta_{\max}])$, and $0$ otherwise. From \cite{ref14}, the distance between the $k$-th GS and $m$-th LEO satellite can be calculated as
\begin{align}
    & d_{k,m,t}  = - R_e \sin \theta_{k,m,t} \nonumber \\
    &+ \sqrt{(R_e \sin \theta_{k,m,t})^2 + (H_{m,t} - R_e)^2 + 2 R_e (H_{m,t} - R_e)},
\end{align}
where $\theta_{k,m,t} \in [\theta_{\min}, \theta_{\max}]$. The channel between the GS and the LEO satellite relies on line-of-sight (LoS) propagation without multipath fading effects \cite{ref16}. However, signal power may be degraded by weather conditions, i.e., rainy or sunny. Considering both free space path loss and rain attenuation, the channel model for GS-to-LEO satellite communication can be expressed as
\begin{align} \label{eq1}
&G(d_{k,m,t}) = ( \frac{ c/f_{12G}}{4 \pi d_{k,m,t}} )^2 Los_{W,t} G_{GS} G_{LEO},
\end{align}
where $c$ is the speed of light and $f_{12G}$ is the carrier frequency in the $12$ GHz band. $Los_{W,t}$ denotes the attenuation caused by weather conditions at time $t$, while  $G_{GS}$ and $G_{LEO}$ represent the antenna gains of the GS and LEO satellites, respectively. The degradation of the received signal strength (RSS) is due to propagation loss and interference. Telecommunication companies and spectrum regulation authorities are now interested in unlocking the $12$ GHz band ($12.2 - 12.7$ GHz) for both terrestrial 5G and non-terrestrial services \cite{ref15}. Sharing the $12$ GHz band could lead to interference due to the coexistence of GS-to-LEO satellite and terrestrial 5G communication. However, our study focuses on the LEO satellite network topology problem and deploys the GSs solely to evaluate the LEO satellite routing path. Therefore, we assume that the GS is located in an area where its communication with the LEO satellite does not face interference issues. The network capacity with interference-free is formulated as 
\begin{equation}
    C_{k,m,t} = \mathcal{B}_{12G}\log_2 (1 + P_{Tx} G(d_{k,m,t}) / \sigma^2),
\end{equation}
where $P_{Tx}$ and $G(d_{k,m,t})$ denote the transmit power and channel gain within the distance $d_{k,m,t}$, respectively. We assume that the transmit power is fixed; thus, it does not vary with time $t$. Additionally, $\mathcal{B}_{12G}$ represents the frequency bandwidth in the 12 GHz band, and $\sigma^2$ is the noise power. Another key performance metric for link selection is the propagation delay, given by $\tau_{k,m,t} = c/d_{k,m,t}$. The optimal link between the $k$-th GS within the visible region and one of the LEO satellites is based on these performance metrics. This is expressed as:
\begin{equation} \label{eq5}
    Link(k,m^*,t)  = \underset{m \in \{1, \dots,M\}}{\mathrm{argmin}}( \{Q(k,m,t) | Q(k,m,t) > 0\}),
\end{equation}
where $Q(k,m,t) = \mathbf{1}_{\{\theta_{k,m,t} \in [\theta_{\min}, \theta_{\max}]\}} (\tau_{k,m,t} + 1/C_{k,m,t})$. Eq.\ref{eq5} indicates the selection of the lowest $Q(k,m,t)$ given $Q(k,m,t) > 0$.

\subsection{Network Model for Inter-Satellite Links}

Unlike the visibility region for GS-to-LEO satellites, two LEO satellites can see each other when their line of sight is not blocked by the atmospheric layer. Therefore, the inter-satellite visibility is determined by the altitude of a triangle, denoted by $\Gamma_{i,j,t}$. In this triangle, the base and the perpendicular sides represent the distances from the Geocenter of Earth to the $i$-th and $j$-th LEO satellites, respectively, while the hypotenuse measures the distance between the two LEO satellites \cite{ref17}. Using the triangle formula, the altitude of a triangle $\Gamma_{i,j,t}$ can be easily calculated as the functions of $H_{i,t}$, $H_{j,t}$, and $D_{i,j,t}$, where $D_{i,j,t}$ is the distance between the $i$-th and the $j$-th LEO satellites. The $i$-th and $j$-th are considered to be visible to each other if their relative altitude $\Gamma_{i,j,t}$ adheres to the following constraint:
\begin{equation} \label{eq6}
    \Gamma_{i,j,t} > \Gamma_{Atmosp},
\end{equation}
where $\Gamma_{Atmosp} = R_e + \gamma_{Atmosp}$. The atmospheric layer height is $50$km $(\gamma_{Atmosp} = 50km)$. The distance between two LEO satellites $D_{i,j}$ can be calculated using Cartesian coordinates in three dimensions, as described in \cite{ref3} (see \cite{ref18} for a more detailed calculation).
When airborne above Earth's atmosphere layer, the communication channel between LEO satellites is modeled as free-space path loss without fading and weather attenuation. The channel power degrades depending on the distance, and it is given by:
\begin{align} \label{eq8}
&G_{LEO}(D_{i,j,t}) = \nonumber \\
& \left(  c/f_{12G} 4 \pi D_{i,j,t} \right)^2 Los_{Pol} Los_{Mis}G_{RX\_LEO} G_{TX\_LEO},
\end{align}
where $Los_{Pol}$ and $ Los_{Mis}$ are the polarization and misalignment antenna losses, respectively. With an interference-free scenario, the mathematical expression for the space network capacity can be written as:
\begin{equation} \label{eq9}
    S_{i,j,t} = \mathcal{B}_{12G} \log_2\left( 1 + SNR_{i,j,t}\right),
\end{equation}
where $SNR_{i,j,t} = P_{Tx} G_{LEO}(D_{i,j,t}) / \sigma^2$ represents the signal-to-noise ratio. The network latency between the two LEO satellites is 
\begin{equation} \label{adq10}
    L_{i,j,t} = \frac{c}{D_{i,j,t}}.
\end{equation}

\subsection{Optimization Problem Formulation}

The purpose of this study is to create an efficient space network topology that maximizes \textit{\textbf{network capacity while minimizing latency and link churn}}. Here, link churn refers to the path changes in satellite networks. Therefore, reducing link churn can minimize communication disruptions, resulting in more stable and continuous service. Given $U$ number of optical link ports at each LEO satellite, usually $4$ in today's LEO satellite constellation~\cite{ref22}), each LEO satellite can connect with any $U$ LEO satellites other satellites within its visibility and propagation coverage area. The space network topology design enables each LEO satellite to select and connect to only the best $U$ LEO satellites from all available options, which optimizes the objective function in a dynamic environment. To achieve this, we formulate an optimization problem where the objective function aims to maximize network capacity while minimizing latency and link churn. This optimization is subject to the following constraints: (C1) the number of link connections allowed for each satellite, (C2) ISL duplex, (C3) inter-satellite visibility, and (C4) propagation distance. Furthermore, the control variable represents the link connection between the satellites. Let $\phi_{i,j,t}$ be a binary variable that denotes the connection status between the $i$-th and $j$-th LEO satellites at time $t$. $\phi_{i,j,t}$ is equal to $1$ if the $i$-th LEO satellite is connected to the $j$-th satellite, and $0$ otherwise. $\Psi_t = \{\phi_{i,j,t} | i = {1,\dots,M}, j = \{1, \dots, M\}\}$ denotes a set of link connections among the LEO satellites at time $t$. The optimization problem is formulated as

\begin{align} \label{eq10}
    &\textbf{P1:} \max_{ \Psi_t } ( \sum_{i = 1}^M \sum_{j = 1 \setminus \{ i\}}^M \phi_{i,j,t} ( S_{i,j,t} + \frac{1}{L_{i,j,t}} +  \phi_{i,j,t-1} ) )  \nonumber \\
    &\text{ \quad s.t.  C1: } \sum_{j=1}^M \phi_{i,j,t} \le U, \forall i \in \{1,\dots,M\} \nonumber \\
    &\text{ \quad \quad \quad C2: } \phi_{i,j,t} = \phi_{j,i,t}, \forall i,j \in \{1,\dots,M\}, \nonumber \\
    &\text{ \quad \quad \quad C3: }  \Gamma_{i,j,t} > \Gamma_{Atmosp}, \forall i,j \in \{1,\dots,M\}, \nonumber \\
    &\text{ \quad \quad \quad C4: } D_{i,j,t} < D_{Max}, \forall i,j \in \{1,\dots,M\},
\end{align}
where $D_{Max}$ represents the maximum distance at which two LEO satellites can communicate with each other. The objective function in (\ref{eq10}) will be zero when the $i$-th satellite is not connected to the $j$-th satellite ($\phi_{i,j,t}=0$), and greater than zero otherwise. Additionally, if the $i$-th satellite is connected to the $j$-th satellite at two consecutive time steps $t-1$ and $t$, an inverse objective function of the link churn will be equal to $1$ ($\phi_{i,j,t}\phi_{i,j,t-1} = 1$), and $0$ otherwise. Therefore, maximizing the objective function in (\ref{eq10}) is equivalent to maximizing network capacity while minimizing latency and link churn.

\section{DTEG-Based Optimal Topology Design (DoTD) Algorithm}

\begin{figure} [t]
    \centering
    \includegraphics[width=0.9\linewidth]{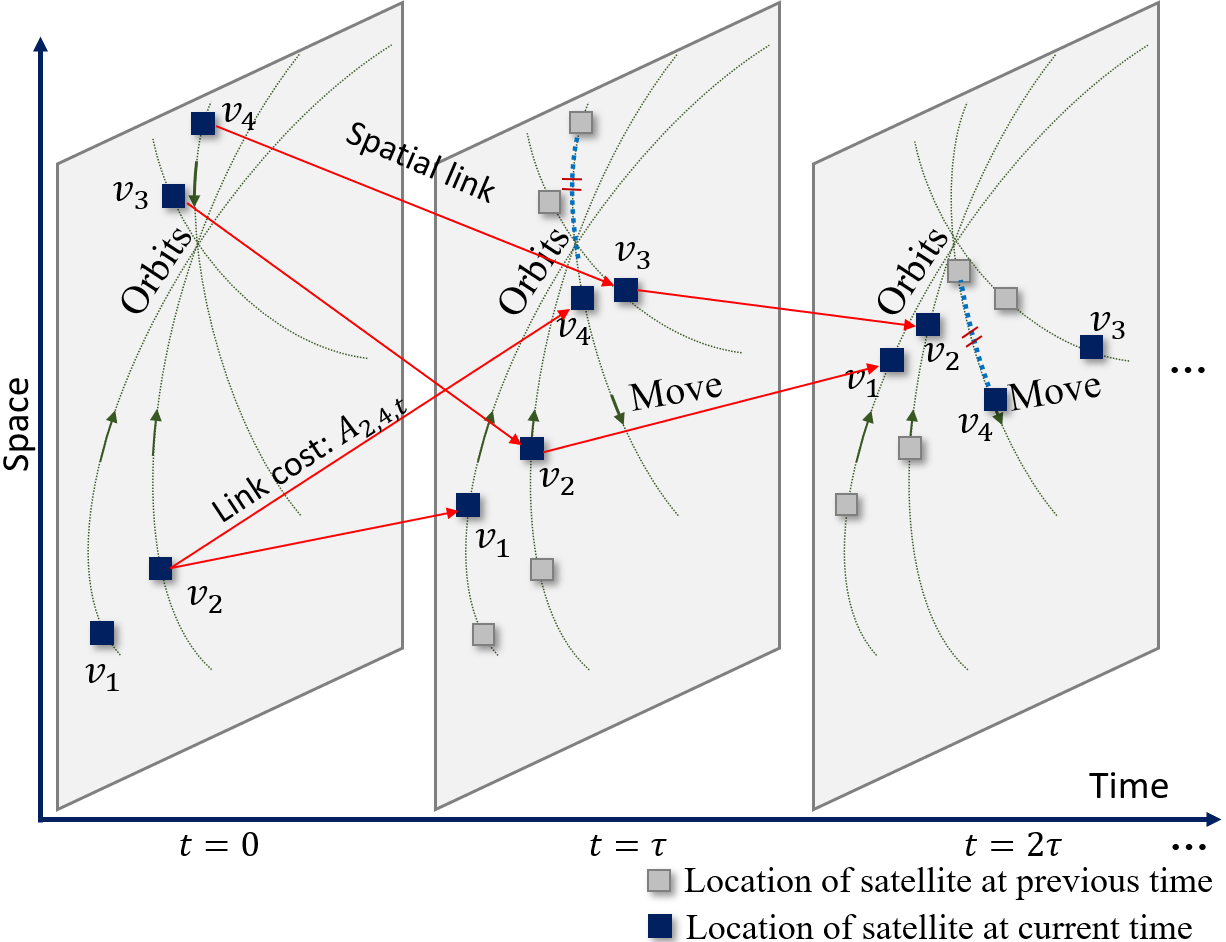}
    \caption{DTEG graph representation.} 
    \label{fig2}
    \vspace{-0.2in}
\end{figure}

To solve the optimization problem, we first construct a Dynamic Time-expanded (DTEG) graph that comprehensively captures the communication opportunities for a given satellite at current time $t$ at future time slots, say, $(t+\tau)$ up to $T$, where $\tau$ is the time period length, and $T$ is the expanded time horizon (say, $10$ minutes). The fixed trajectories of LEO satellites -- including constant velocity of satellite, orbital period, and altitudes together with Earth's steady rotation -- facilitate accurate predictions of their future achievable performances even in the dynamic movement of space network environment (see in Fig.\ref{fig2}). Next, we introduce a novel DTEG-based Topology Design (DoTD) algorithm that maximizes the objective function of problem \textbf{P1} (over an expanded time horizon $T$). The proposed DoTD algorithm is a dynamic programming based algorithm and runs in polynomial time.


\subsection{DTEG Construction}
Consider a total time duration $T$ to be divided into discrete and equal time slots (of interval length $\tau$), such as, $\{0, 1, \dots, T\}$. At a given time slot $t$, let $G_t = \{V, E_t\}$ be an undirected graph that represents the snapshot of space network, where $V$ is the set of nodes, and $E_t \subseteq (V \times V)$ is the set of communication links between node-pairs at time slot $t$. For most scenarios and ease of presentation, we assume $V$ is constant over time, where the set of potential links may change due to the varying satellite node positions, and varying channel conditions. This series of graph snapshots $\{G|t = 1 \dots T\}$ is used for the construction of DTEG graph $\mathcal{G} = (\mathcal{V}, \mathcal{E})$. Primarily, the DTEG graph $\mathcal{G}$ is a layered graph, where each layer corresponds to a discrete time interval of length $t$. 

\textbf{Node Structure:} Given the series of graph snapshots $\{G_t|t = 1 \dots T\}$, the DTEG graph has a total of $(\hat{T} +1)$ layers of nodes, where $\hat{T} = T/\tau$. Thus, the node set $\mathcal{V}$ is: $\mathcal{V} = \{v_{i,t}|i \in \{0, \dots, |V|\}, \text{and } t \in [0, \hat{T}]\}$. As a result, the total number of nodes is $|V|\times (\hat{T}+1)$. 

\textbf{Link Structure:} DTEG graph $\mathcal{G}$ will include spatial links only. A spatial link represents a communication possibility from a satellite node $v_i$ to another node $v_j$ at time $t$. Formally, a spatial link is a tuple $(v_{i,t}, v_{j,t+1}$), where $i \neq j$, and the link conditions are met.

\textbf{Link Score Cost:} We define the cost $A_{i,j,t}$ of a link $e \in \mathcal{E}$ as its score value. The score value is a weighted sum of normalized values for capacity, latency and link churns, and is discussed in detail in the next subsection. The detail calculation of the score value function is provided in Eq.\ref{eq13}.

\subsection{Algorithm Description}

Leveraging the DTEG graph (representing dynamic space network constellation), we propose a space network topology design called the DoTD algorithm that is based on dynamic programming principle \cite{ref23}. The proposed DoTD algorithm tackles the optimization problem in (\ref{eq10}) by transforming it into a \textbf{score}-based optimization problem. \textit{The concept of a score is widely understood as an indicator used to evaluate student outcomes. Top students are ranked based on the assessment of their time-dependent scores obtained from exams, assignments, and homework.} With this key idea, the proposed DoTD algorithm enables each LEO satellite to evaluate others based on their scores and select the top $U$ satellites with the highest scores to establish link connections. The objective function in problem \textbf{P1} is transformed into the score-based objective function as follows. Due to the differing scales of network capacity ($S_{i,j,t}$), latency ($L_{i,j,t}$), and link churn ($\phi_{i,j,t}$), the algorithm may focus on the metric with the highest scale during selection. This issue can be addressed by normalizing the KPMs to a common range from $0$ to $1$. Let $\Bar{S}_{i,j,t}$, $\Bar{L}_{i,j,t}$, and $\Bar{\phi}_{i,j,t}$ be the normalized values for capacity, latency, and link churn, respectively. They can be calculated as follows: 
\begin{align} \label{eq11}
    \Bar{S}_{i,j,t}  = \frac{S_{i,j,t}}{S_{\text{Max},t}}; \Bar{L}_{i,j,t}  = \frac{L_{i,j,t}}{L_{\text{Max},t}}; \Bar{\phi}_{i,j,t}  = \frac{\phi_{i,j,t}}{U},
\end{align}
where $S_{\text{Max},t}$ and $L_{\text{Max},t}$ are the maximum values of link capacity and latency at time $t$, respectively. The update rule for maximum value is given by
\begin{align} \label{eq12}
    &g_{\text{Max},t} = \max \{g_{\text{Max},t-1}, \nonumber\\ 
    & \max_{i\in\mathcal{M},j\in\mathcal{M}}(\textbf{1}_{\{\Gamma_{i,j,t} > \Gamma_{Atmosp}\}} \textbf{1}_{\{D_{i,j,t} < D_{Max}\}} g_{i,j,t})\},
\end{align}
where $g_{\text{Max},t} \in \{ S_{\text{Max},t}, L_{\text{Max},t} \} $, and $\mathcal{M}$ is the set of satellites. This time-dependent maximization allows the calculation of the maximum values for capacity and latency, ensuring they are always greater than the values at any given time. Normalizing using these maximum values maintains the scale of achievable capacity and latency over different time steps. The link cost, which normalizes the objective function in \textbf{P1}, can be written as:
\begin{align} \label{eq13}
    & A_{i,j,t} = w_1 \Bar{S}_{i,j,t}  + w_2(1 - \Bar{L}_{i,j,t}) + (1 - w_1 - w_2) \Bar{\phi}_{i,j,t-1},
\end{align}
where $w\in \{w_1 , w_2, 1-w_1-w_2\}$ are the weights of the objective function. The constraints on the weights are: $w_1 \in [0,1], w_2 \in [0,1]$, and $w_1+w_2 \le 1$. With this design, maximizing the simplified function $A_{i,j,t}$ is equivalent to maximizing the objective function in \textbf{P1}. The score value function is given by: 
\begin{equation} \label{eq14}
    \alpha_{i,j,t} = \textbf{1}_{\{\Gamma_{i,j,t} > \Gamma_{Atmosp}\}} \textbf{1}_{\{D_{i,j,t} < D_{Max}\}}(A_{i,j,t} + \Pi_{j,t-1}), 
\end{equation}
where $\Pi_{j,t-1}$ is the historical score function of the $j$-th node at time $t-1$, and it is set to zero at initial time $(\Pi_{j,t=0} = 0)$. The optimal link can be obtained as follows:
\begin{equation} \label{eq15}
    \phi_{i,j,t}^* = \underset{l \in \mathcal{M}}{\mathrm{argmin}} ( \{ \alpha_{i,l,t} | (\sum_{k=1}^M \phi_{i,k,t}^* < U \cap \sum_{k=1}^M \phi_{j,k,t}^* < U ) \} ).
\end{equation}
The design of the score function in (\ref{eq14}) ensures that the $i$-th satellite can select any $j$-th satellite within the inter-satellite visibility and communication range, thus satisfying constraints C3 and C4. The optimal link selection in (\ref{eq15}) allows the $i$-th satellite to choose the $j$-th satellite with the highest score given that its total link connections are less than $U$ and the connections of the $j$-th satellite are also less than $U$, thus adhering to constraints C1 and C2. The binary variable representing the $j$-th satellite connected to the $i$-th satellite is given by $\phi_{j,i,t}^* = \phi_{i,j,t}^*$. Therefore, the optimization problem \textbf{P1} was solved. The update rule for the score function is:

\begin{equation} \label{eq16}
    \Pi_{i,t} = \frac{1}{U} \sum_{j = 1 \setminus \{ i\}}^M \phi_{i,j,t}^* \left( A_{i,j,t} + \Pi_{j,t-1} \right).
\end{equation}

\begin{lem}
The time-dependent maximum value \( g_{\text{Max},t} \) defined in Eq. \ref{eq12} is the achievable maximum capacity, latency, and link churn, denoted by $g$, up to time $t$.
\end{lem}

\begin{proof}
Assume that the initial value \( g_{\text{Max},t=0} \) is given at time \( t = 0 \), which represents the maximum value of \( g \). At time \( t-1 \), \( g_{\text{Max},t-1} \) correctly represents the maximum value of \( g \) for all \( i, j \in \mathcal{M} \) up to time \( t-1 \). Suppose there exists another value of \( g' \), such that \( g' > g_{\text{Max},t} \). To demonstrate that $g_{Max,t}$ is the maximum value up to time $t$, we prove that the assumption $g' > g_{\text{Max},t}$ is not true. Under this assumption, there are two possible cases to consider for \( g_{\text{Max},t} \). \\
$\bullet$ \textbf{Case 1}: Assume that \( g_{\text{Max},t} = g_{\text{Max},t-1} \). By the inductive hypothesis, \( g_{\text{Max},t-1} \) is the maximum value of \( g \) up to time \( t-1 \). Additionally, if \( g_{\text{Max},t} = g_{\text{Max},t-1} \), then \( g_{\text{Max},t} \) must be at least as large as any \( g_{i,j,k} \) for \( 0 \le k \le t-1 \). This implies \( g_{\text{Max},t-1} \geq g' \), which contradicts our assumption that \( g' > g_{\text{Max},t} \). \\
$\bullet$ \textbf{Case 2}: From Eq.\ref{eq12}, \( g_{\text{Max},t} \) is the maximum value of \( g_{i,j,t} \) at time \( t \) under the constraints \( \Gamma_{i,j,t} > \Gamma_{Atmosp} \) and \( D_{i,j,t} < D_{Max} \). If there exists some \( g' \) such that \( g' > g_{\text{Max},t} \), then \( g' \) should have been included in the set of values for \( g_{i,j,t} \), which also contradicts the definition of \( g_{\text{Max},t} \). From both cases, the assumption that \( g' > g_{\text{Max},t} \) is contradiction. Therefore, the value \( g_{\text{Max},t} \) is indeed the maximum of \( g_{\text{Max},t-1} \) and \( \max_{i \in \mathcal{M}, j \in \mathcal{M}} ( \mathbf{1}_{\{\Gamma_{i,j,t} > \Gamma_{Atmosp}\}} \mathbf{1}_{\{D_{i,j,t} < D_{Max}\}} g_{i,j,t} ) \).
\end{proof}

\begin{lem}
The update rule of the score function in (Eq. \ref{eq16}) remains finite even as the observation time approaches infinity: $\Pi_{i,t} < \infty, T \to \infty$.
\end{lem}
This ensures that the proposed DoTD algorithm can continuously evaluate the scores of satellites and select connections at all time steps.

\begin{proof}
The score of the LEO satellite is equal to zero at the initial time step ($\Pi_{i,t=0} = 0$). Thus, the scores from time $t=1$ to $t=T$ are given by:
\begin{align} \label{eq17}
    &\Pi_{i,1} = \frac{1}{U} \sum_{j = 1 \setminus \{ i\}}^M \phi_{i,j,1}^*  A_{i,j,1} \nonumber \\
    &\Pi_{i,2} = \nonumber \\
    &\frac{1}{U} \sum_{j = 1 \setminus \{ i\}}^M \phi_{i,j,1}^*  A_{i,j,1} +\frac{1}{U^2} \sum_{j = 1 \setminus \{ i\}}^M \sum_{k = 1 \setminus \{ j\}}^M  \phi_{i,j,1}^*\phi_{j,k,2}^*  A_{j,k,2}  \nonumber \\
    &\vdots \nonumber \\
    &\Pi_{i,T} = \nonumber \\
    &\frac{1}{U} \sum_{j = 1 \setminus \{ i\}}^M \phi_{i,j,1}^*  A_{i,j,1} +\frac{1}{U^2} \sum_{j = 1 \setminus \{ i\}}^M \sum_{k = 1 \setminus \{ j\}}^M  \phi_{i,j,1}^*\phi_{j,k,2}^*  A_{j,k,2}   \nonumber \\ 
    &+ \dots + \frac{1}{U^T} \underbrace{ \sum_{j = 1 \setminus \{ i\}}^M \dots \sum_{k = 1 \setminus \{ l\}}^M}_T ( \underbrace{\phi_{i,j,1}^* \dots \phi_{l,k,T}^*}_T  A_{j,k,T}  ) 
\end{align}
From (\ref{eq13}), $\Bar{L}_{i,j,t} > 0$ and $\Bar{\phi}_{i,j,t-1} < 1$ $\Rightarrow$ $A_{i,j,t}< 1$, $\forall t$. Then, from (\ref{eq17}), if $T \to \infty $ $\Rightarrow$ $U^T \to \infty$; thus, $\Pi_{i,T} < \infty$.
\end{proof} 

\begin{algorithm} [t]
\small
\caption{DoTD Algorithm}\label{alg1}
\begin{algorithmic}[1]
\State \textbf{Input:} 
\State \textit{TLE\_Data\_File in \cite{ref9}}: File containing TLE data 
\State \textit{Set\_Start\_Time}: Start time for the experiment 
\State \textit{Set\_End\_Time}: End time for the experiment
\Statex---------------------\textbf{Algorithm}------------------------------------
\State \textit{Initialization}: Set maximum capacity $S_{i,j,t}$, latency $L_{i,j,t}$, and score value function $\Pi_{i,j,t=0}$ to zero
\For{$t = 1 \text{ to } T$}
    \State Read \textit{TLE\_Data} to generate LEO satellite constellation \cite{ref19}
    \State Update time and generate locations of LEO satellites \cite{ref19}
    \State Compute the visible altitude $\Gamma_{i,j,t}$ using Eq.\ref{eq6}
    \State Generate the distances between LEO satellites $D_{i,j,t}$ \cite{ref19}
    \State Determine the network capacity $S_{i,j,t}$ and latency $L_{i,j,t}$ using
    \StateX Eq.\ref{eq9} and Eq.\ref{adq10}, respectively.
    \State Update the maximum capacity $S_{Max,t}$ and latency $L_{Max,t}$
    \StateX using Eq.\ref{eq12}
    \State Normalize the capacity, latency, and link churn using Eq.\ref{eq11}
    \State Obtain the normalized objective function $A_{i,j,t}$ using Eq.\ref{eq13}
    \State Compute the score-based normalized objective function $\alpha_{i,j,t}$ 
    \StateX using Eq.\ref{eq14}
    \For{$ \text{each satellite } i = 1 \text{ to } M$}
        \For{$j = 1 \text{ to } M$}
            \If{$\sum_{k=1}^M \phi_{i,k,t}^* < U \text{ and } \sum_{k=1}^M \phi_{j,k,t}^* < U$} 
            \State Select the optimal link: $\phi_{i,j,t}^* = \underset{l \in \mathcal{M}}{\mathrm{argmin}}(\alpha_{i,l,t})$
            \State Update ISL duplex: $\phi_{j,i,t}^* = \phi_{i,j,t}^*$
            \EndIf 
        \EndFor
    \EndFor
    \State Update the score function:
    \Statex \quad \quad \quad \quad $\Pi_{i,t} = \frac{1}{U} \sum_{j = 1 \setminus \{ i\}}^M \phi_{i,j,t}^* \left( A_{i,j,t} + \Pi_{j,t-1} \right)$
\EndFor
\Statex----------------------------------------------------------------------
\State \textbf{Output:} Optimal link $\phi_{i,j,t}^*$ to create the best topology
\end{algorithmic}
\vspace{-0.2cm}
\end{algorithm}
The proposed DoTD algorithm is summarized in \textbf{Algorithm \ref{alg1}}. This algorithm allows each LEO satellite to select the best $U$ satellite for connection to create an optimal dynamic time-expanded graph network topology.

\textbf{\underline{Time complexity:}} The proposed DoTD algorithm pre-computes the achievable performance metrics—such as capacity, latency, and link churn—for all communication links between the satellite-pairs over $T$ timestamps, in order to evaluate the score value function given in Eq.\ref{eq16}. In this context, the time complexity of the proposed algorithm equals to $\mathcal{O}(\hat{T}M^2)$, where $\hat{T} = T/\tau$. Here, $T$ represents the total time horizon and $\tau$ is time period length. $T$ can range from a few minutes (e.g., 10 minutes) up to the maximum duration required for one complete Earth rotation (approximately $90–100$ minutes). For our DoTD algorithmic purposes, we consider $T$ to be on the order of few ($10$'s of) minutes. This is because satellite positions are accurately predictable over shorter time frames, and there is no practical need to forecast satellite positions several minutes in advance. Additionally, $\tau$ is expected to be on the order of seconds (e.g., 1 second) or a few hundred milliseconds. Given this, $\hat{T}$ can be treated as a constant and thus ignored. Consequently, the time complexity of the proposed algorithm simplifies to $\mathcal{O}(M^2)$.


\noindent \textbf{\textit{Open Shortest Path First (OSPF) Routing Algorithm:}} 
We apply the Open Shortest Path First (OSPF)-based method to the optimal network topology obtained by our proposed DoTD algorithm. The OSPF-based DoTD routing path is introduced to an optimal routing path from a GS source to its destination. The network model for the GS-to-LEO satellite in sub-section \ref{subIIB} provides a solution that allows the GS to connect with the visible satellite (see Eq. \ref{eq5}). The proposed OSPF-based DoTD routing algorithm optimizes the routing path from the selected satellite to the satellite connected to the destination GS. Our approach works by exchanging information about the IP addresses and achievable scores within optimal network topology created by DoTD in \textbf{Algorithm \ref{alg1}}. This information enables the source satellite to find its destination and compute the aggregated scores, allowing it to select a routing path that maximizes capacity and minimizes latency. 
The proposed \textbf{Algorithm \ref{alg1}} updates the network topology at intervals of $T$, where $T$ can be several minutes or more (e.g., 10 minutes). This update interval is restricted by constraints related to configuration complexity, service continuity, and stability, which prevent updates at sub-second intervals. To manage this, the DoTD algorithm divides the total time $T$ into multiple time slots with a granularity of $\tau$ (e.g., $\tau=1s$), which corresponds to the communication time between two LEO satellites. At any time $t$, the algorithm pre-computes network performance for each time slot $\tau$ from the current time $t$ to the future time $t+T$, and aggregates the score value functions of all satellites. This approach enables each satellite to select $U$ satellites for establishing the connections that optimize performance at each timestamp from $t$ to $t+T$. When a source node requests communication with a destination, the OSPF algorithm is applied to optimize the routing path within the topology established by \textbf{Algorithm \ref{alg1}}. The OSPF algorithm leverages the network topology derived from the DoTD to determine the optimal routing whenever a communication request is made. 

\begin{table}[!t]
\caption{Network Parameters} \label{Tab1}
\centering
\begin{tabular}{p{5.5cm}p{2.5cm}}
\hline \hline
    Parameter Settings & Values \\
    \hline
    Speed of light $c$ & $3\times 10^8$ m/s\\
    Speed of LEO satellite  & $7.66$ km/s \\
    Atmospheric layer height $\gamma_{Atmosp}$ & $50$ Km\\
    Earth's radius $R_e$ &  $6378$ Km \\
    Carrier frequency $f_{12G}$ & $12.2$ GHz \cite{ref20}\\
    Total bandwidth $\mathcal{B}_{12G}$ & $100$ MHz \cite{ref20}\\
    Polarization loss $Los_{Pol}$ & $4.5$ dB \\
    Misalignment loss $Los_{Mis}$ & $0.5$ dB\\
    GS antenna gain $G_{GS}$ & $33.2$ dBi\\
    LEO satellite antenna gain $G_{LEO}$ & 40 dBi\\
    Maximum communication range $D_{Max}$ & 7000 Km \cite{ref21}\\
    Number of link selections $U$ & 4 \cite{ref22}\\
    Number of LEO satellites & 907 \\
    Weights of objective functions $w_1 = w_2$  & 0.4 \\ 
\hline 
\end{tabular}
\end{table}

\section{Experimental Results}

\subsection{Experimental Setup}

\begin{figure}[t]
    \centering
    \includegraphics[width=8.5cm]{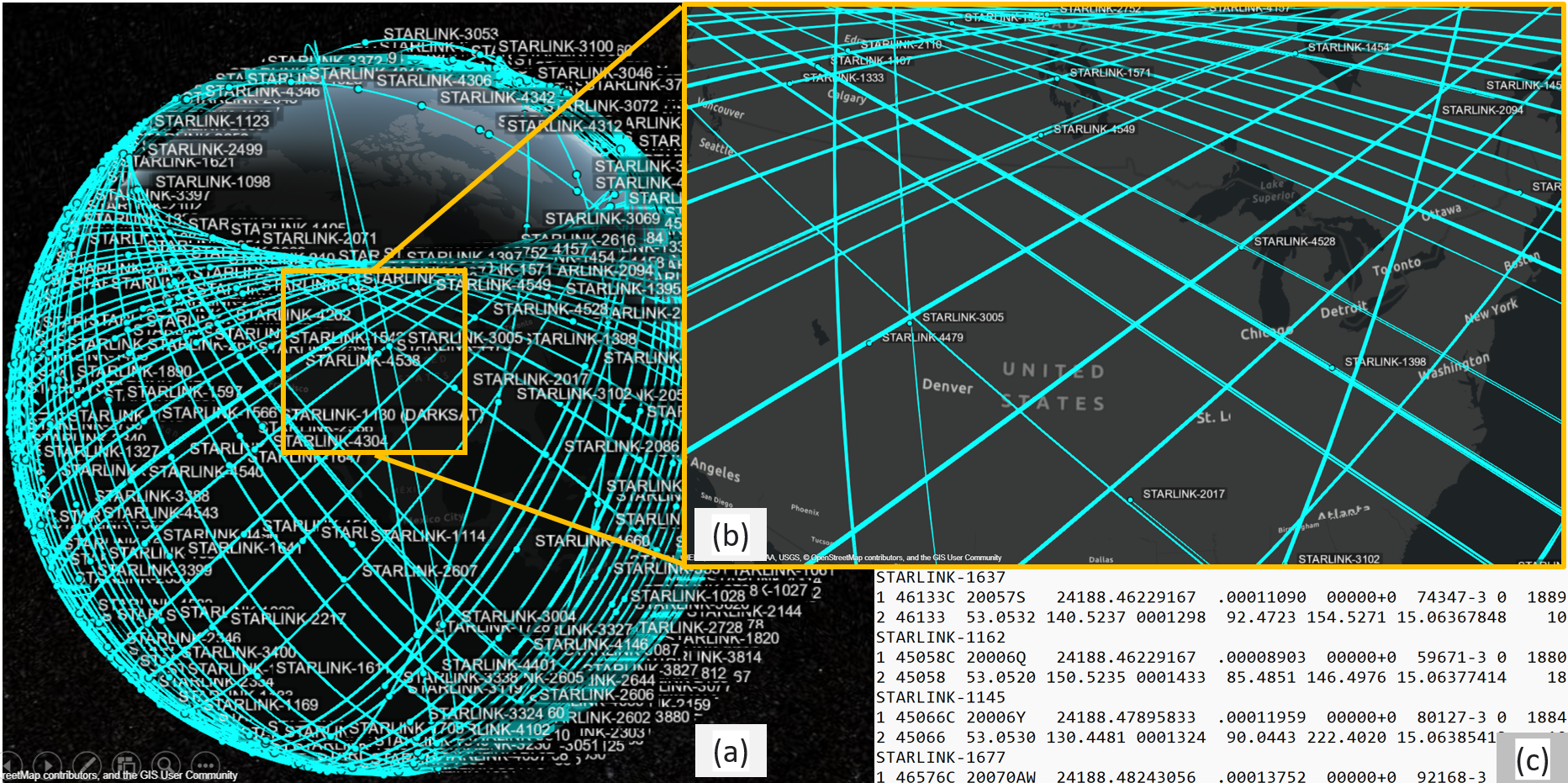}
    \caption{SpaceX's Starlink LEO satellites constellation}
    \label{fig3}
    \vspace{-0.2in}
\end{figure}

\begin{table*}[!t]
\caption{GS deployments on Earth using GCS} \label{Tab2}
\vspace{-0.1in}
\centering
\begin{tabular}{p{2cm}p{2cm}p{2cm}p{2.5cm}p{2cm}p{2cm}p{2cm}}
\hline \hline
    Scenarios  & \textbf{\underline{Sources}}  & Latitudes  & Longitudes &  \textbf{\underline{Destinations}}  & Latitudes  & Longitudes \\ 
    \hline 
    S1 & Sydney  & $-33.865143$ & $151.209900$ &  Darwin  & $-12.46$ & $130.84$ \\
    S2 & Miami  & $25.761681$ & $-80.191788$ & Calgary & $51.049999$ & $-114.066666$ \\
    S3 & New York  & $40.730610$ & $-73.935242$ & Miami  & $25.761681$ & $-80.191788$ \\
    S4 & New York  & $40.730610$ & $-73.935242$ & San Francisco & $37.773972$ & $-122.431297$ \\
    S5 & Phnom Penh  & $11.562108$ & $104.888535$ & Kathmandu  & $27.700769$ & $85.300140$ \\
\hline 
\end{tabular}
\vspace{-0.2in}
\end{table*}

CelesTrak in \cite{ref9} has received permission from SpaceX to provide live updates of TLE datasets, which are used to generate the real LEO satellite constellation emulator. These general perturbation (GP) datasets are derived from radar and optical observations conducted by the US Space Surveillance Network (SSN). Additionally, CelesTrak provides test cases that demonstrate the high accuracy of its localization predictions. For this experiment, our team collected TLE datasets on June 10, 2024, at 8:23:00 UTC and analyzed them using Aerospace and Satellite Communications Toolboxes in \cite{ref19} to create a real-time LEO satellite network emulator that perfectly synchronizes to the real-world LEO satellites, as illustrated in Fig.\ref{fig3}. The accuracy of location estimations for LEO satellites, obtained using CelesTrak, and Aerospace and Satellite Communications Toolboxes, is facilitated by the constant speed of the satellites, the fixed altitude of their orbital planes, and Earth's steady rotation. GSs are deployed in different states or countries based on a geographic coordinate system (GCS), which measures positions on Earth using latitude and longitude. Our proposed DoTD approach in \textbf{Algorithm \ref{alg1}} generates the network topology that allows each satellite to connect with four other satellites. This limited number of link connections follows a form of the $+$Grid structure where each satellite is linked to its $4$ neighbors. The capacity and latency of each link are then calculated using the network parameters given in TABLE \ref{Tab1}. All this information is fed into \simulator~\cite{kassem2024xeoverse} to generate routing paths from the source to the destination using OSPF. 

\begin{figure}[t]
    \centering
    \includegraphics[width=7.5cm]{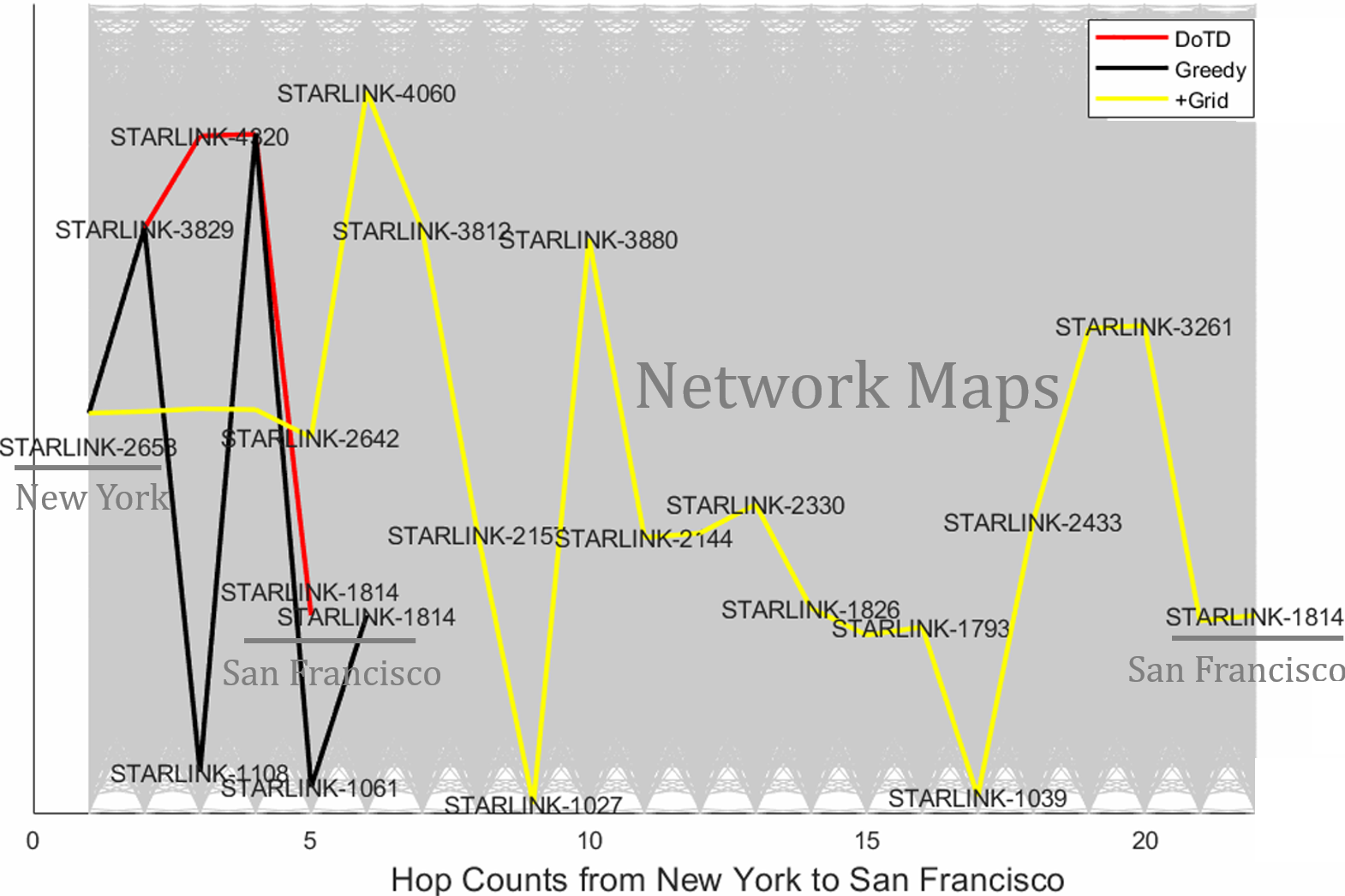}
    \caption{Network maps created by DTEG, Greedy, and $+$Grid: Packets traveling from New York to San Francisco.}
    \label{fig4}
    \vspace{-0.2in}
\end{figure}

\subsection{Baseline Methods}
We evaluate two baseline methods—Greedy and $+$Grid—and compare them with our proposed algorithm. 

$\bullet$ \textbf{Greedy:} The greedy method is the first order of the proposed DoTD algorithm, which establishes $4$ links within its inter-satellite visibility and communication range based on the score value function in Eq.\ref{eq13}. Unlike DoTD, this method captures the score value of each link at the current timestamp and selects the $4$ links with the highest scores compared to other available links to create the space network topology. 

$\bullet$ \textbf{$+$Grid:} This method focuses on
selecting links based on a grid pattern: two LEO satellites are chosen from inter-orbital links, and two are chosen from intra-orbital link. The orbital parameters of the satellites—such as argument of perigee, true anomaly, RAAN, and inclination—determine whether they belong to the same orbital link or not. Each satellite compares its RAAN and inclination with those of others to identify satellites moving in consecutive orbits, and then establishes connections accordingly. 

\subsection{Experimental Results}

\begin{figure*}[h!]
    \centering
    \subfloat{\includegraphics[width=5.3cm]{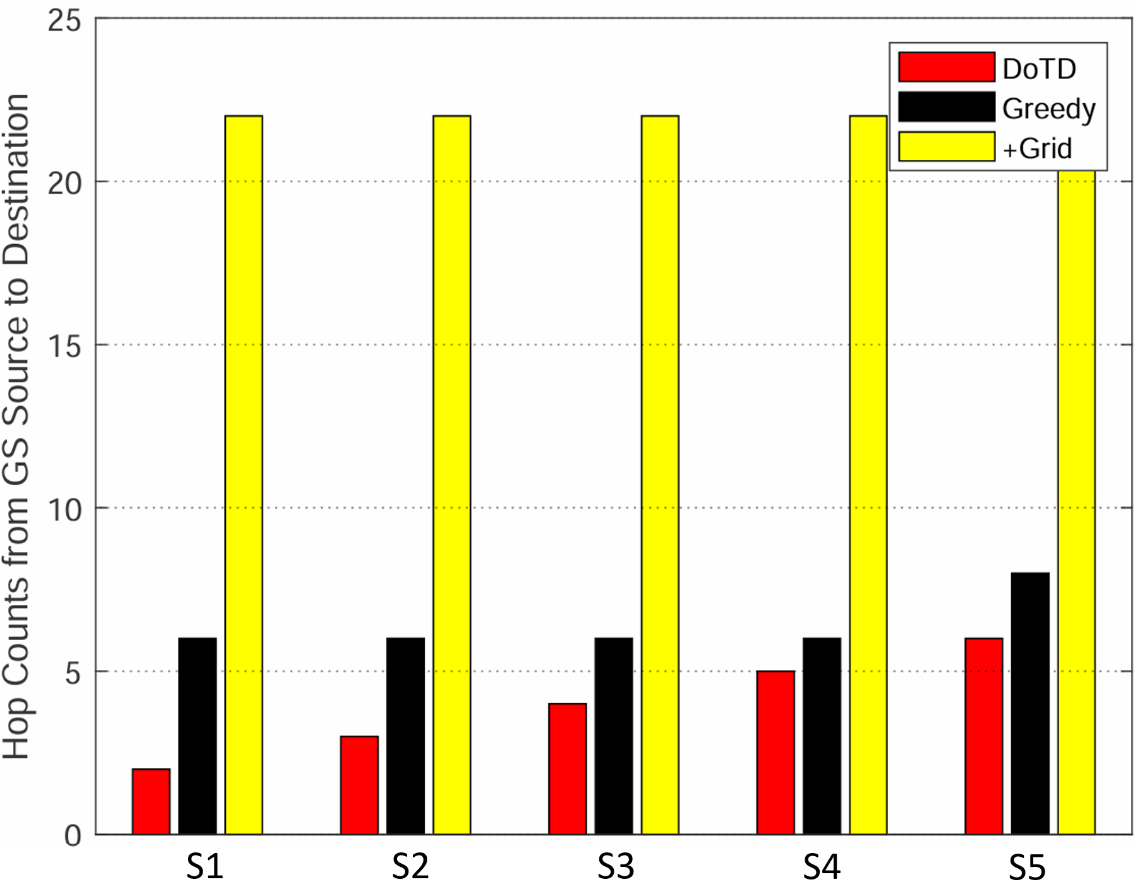}}%
    \qquad
    \subfloat{\includegraphics[width=5.3cm]{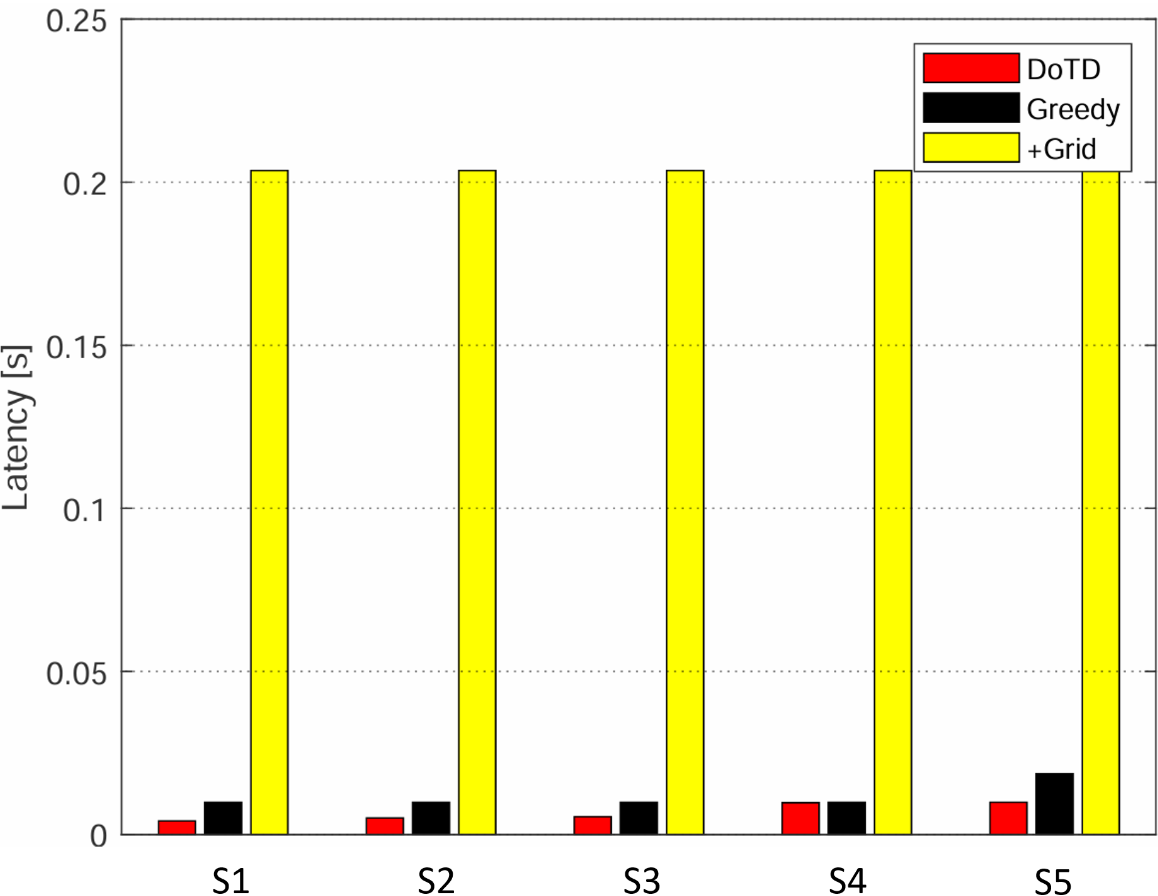}}%
    \qquad
    \subfloat{\includegraphics[width=5.3cm]{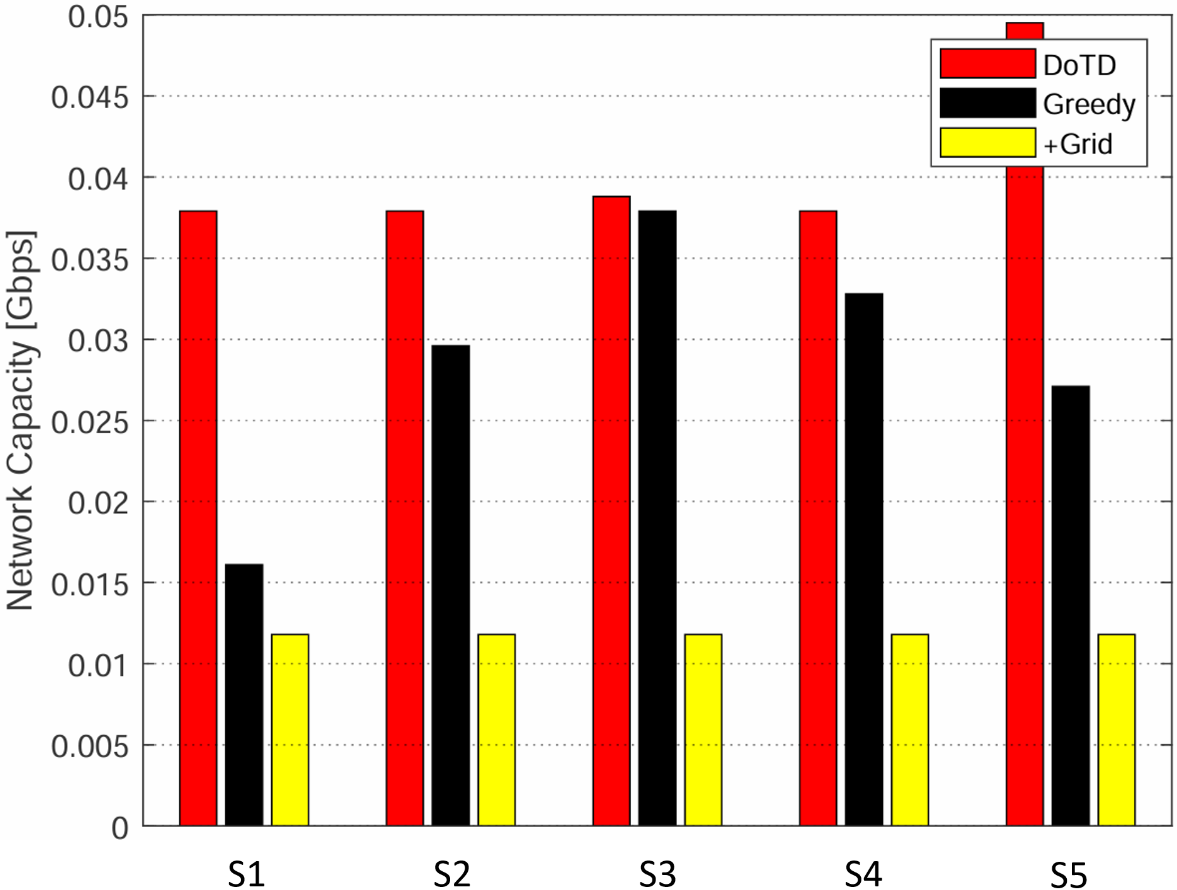}}%
    \vspace{-0.05in}
    \caption{(a) Hop counts from the GS source to the destination, (b) Latency evaluation, (c) Average achievable capacity of each link, under five different scenarios in TABLE \ref{Tab2}.}%
    \label{fig:fig5_fig6_fig7}%
    \vspace{-0.1in}
\end{figure*}

\textit{\underline{Experiment 1:}} The GS source is deployed in New York, while the destination is located in San Francisco, as illustrated in Fig.\ref{fig4}. A network map is created to clearly show the optimal routing paths of the proposed DoTD algorithm (red line) and other two baseline methods: greedy (black line) and $+$Grid (yellow line). This map also illustrates the ISL connections between consecutive LEO satellites, including their IDs, from the GS source to the destination. On June 10, 2024, at 8:23:00, the Starlink-2658 satellite, which is airborne above New York, establishes a connection with the GS source in that state to relay packets to the destination in San Francisco, which is connected to the LEO satellite Starlink-1814. The results demonstrate that the proposed DoTD algorithm outperforms the baseline methods by requiring only $5$ LEO satellites to forward packets to the GS destination across different states. In comparison, the greedy method needs $6$ hops, while the $+$Grid method requires up to $22$ hops. This confirms our earlier discussion (see RCM in Section \ref{sec1}) about the challenges faced by the $+$Grid method due to the imperfect LEO satellite constellation. Faced with this issue together with the satellites' mobility, the $+$Grid method selects $4$ satellites from intra-orbital and consecutive inter-orbital links, which do not align with the optimal routing path. At any time, the satellites in other orbits can be closer to each other than those in the same or consecutive adjunction orbits, which degrades the performance of the $+$Grid method. The updated versions of $+$Grid, namely $\times$Grid and Motif, also emphasize connections to neighboring satellites. Like $+$Grid, these two algorithms experience performance degradation due to imperfections in the satellite constellation (see the real satellite emulator in Fig.\ref{fig3}). The satellites and orbital planes are not evenly spaced. Given these limitations, we ensure that our proposed DoTD algorithm also outperform these two baseline methods. Therefore, we do not present performance evaluations for $\times$Grid and Motif in comparison to our proposed algorithm. In comparison to Greedy, DoTD's ability to pre-estimate the future performance of links based on satellite mobility patterns enables it to determine which satellites should be connected at specific future times to continuously maximize performance. This capability allows DoTD to achieve better performance compared to the greedy method, which only considers the performance of links at the current time for selection.

\textit{\underline{Experiment 2:}} The multiple GS sources and their destinations are deployed across various states and countries on Earth using the GCS, as detailed in TABLE \ref{Tab2}. The network maps are generated using DoTD, greedy, and $+$Grid with five different GS source-destination pairs to measure performance metrics, including hop counts (Fig.\ref{fig:fig5_fig6_fig7}(a)), latencies (Fig.\ref{fig:fig5_fig6_fig7}(b)), and network capacities (Fig.\ref{fig:fig5_fig6_fig7}(c)). These results demonstrate that the proposed algorithm consistently achieves the optimal routing path across all GS deployments on Earth. Applying the DoTD algorithm to optimize the space network topology can reduce the average hop count rate by 10.91\% compared to the greedy method and by up to 81.82\% compared to the $+$Grid method, as illustrated in Fig.\ref{fig:fig5_fig6_fig7}(a). In most cases, reducing the hop count often shortens the path from the GS source to the destination, thereby decreasing the total latency required to send a data packet through the space network topology to its destination. In this experiment, the proposed algorithm achieves a very low latency of only $4.2$ ms, while the greedy method requires $9.9$ ms and the $+$Grid method incurs a much higher latency of up to 203.6 ms for transmitting packets from the same GS source (Sydney) to the destination (Darwin), as shown in Fig.\ref{fig:fig5_fig6_fig7}(b). Additionally, in the evaluation across these five different scenarios (S1-S5), the DoTD algorithm significantly improves network capacity. It achieves an average increase of 28.09\% over the greedy method and up to 70.47\% over the $+$Grid method, as illustrated in Fig.\ref{fig:fig5_fig6_fig7}(c).

\begin{figure}[!t]
    \centering
    \includegraphics[width=6.5cm]{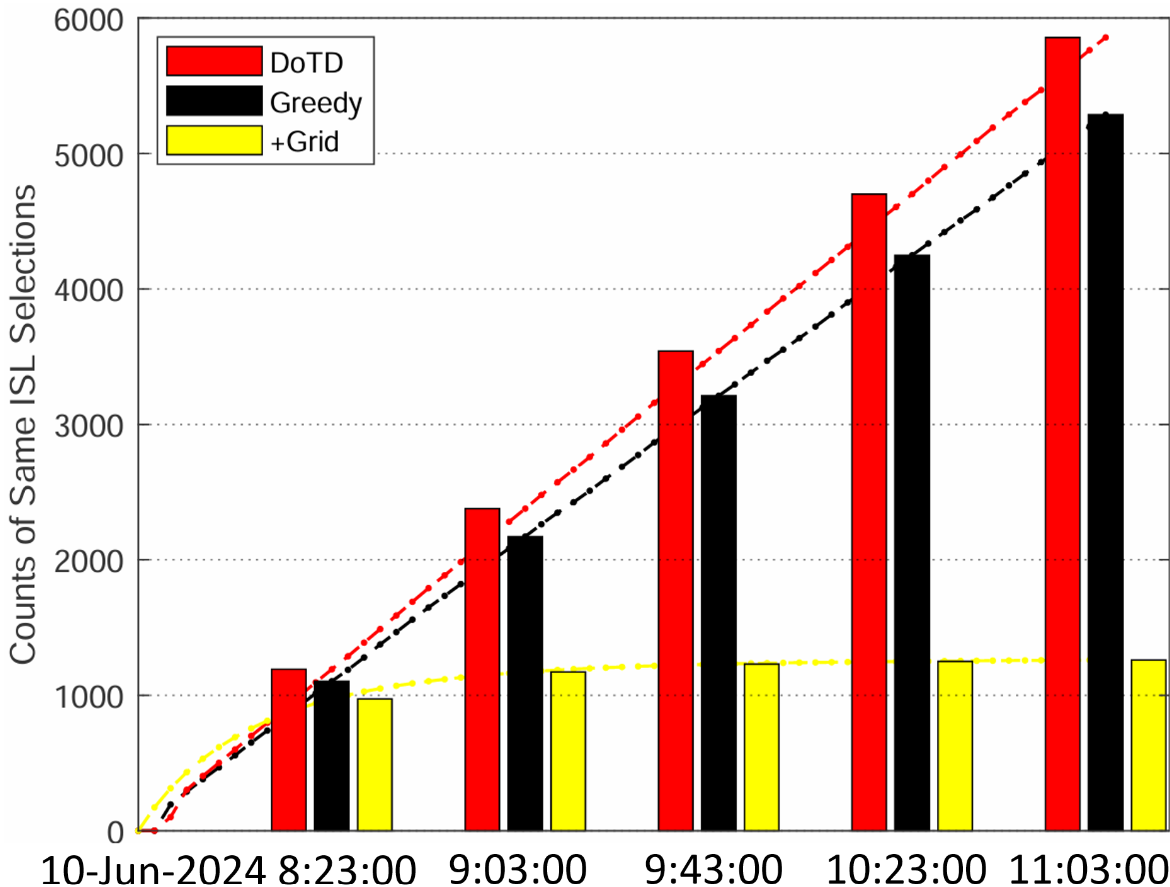}
    \caption{Counts the number of ISL selections that are the same.}
    \label{fig8}
    \vspace{-0.2in}
\end{figure}

\textit{\underline{Experiment 3:}} This experiment evaluates the link churn achieved by the proposed method and compared it to other two baselines. We monitor the changes in ISL connections over a period of 3h and 20 min, from 08:23 on June 10, 2024, to 11:03 on the same day, as shown in Fig.\ref{fig8}. We count the ISL connections of all LEO satellites that connect to the same satellites at two consecutive time steps, which allows for the update of the space network topology. The results reveal that the proposed method maintains the same ISL connections more effectively than the greedy and $+$Grid methods. This improvement is achieved by incorporating link churn into the objective function of optimization problem \textbf{P1}. Thus, DoTD takes link churn into account when selecting satellites that minimize changes in ISL connections. Reducing these changes lowers network re-configuration costs and enhances network stability and service continuity.

\section{Conclusion and Future work}

In this paper, we propose an optimal DoTD algorithm to address the network topology challenges encountered by the neighbor-grid method due to imperfections in LEO satellite constellations. Our approach focuses on selecting ISL connections that consistently achieve optimal performance metrics—such as network capacity, latency, and link churn—over time. To this end, we formulated a score value function by normalizing multiple objective functions to a common range from $0$ to $1$. We then introduce the DoTD algorithm, which pre-estimates the achievable score value function for each satellite at future timestamps and selects the optimal satellites for establishing connections based on these predictions. By evaluating and selecting links using this score value function, the proposed approach ensures that the network topology remains optimized for maximum performance at all times. Our experiments with the Starlink LEO satellite constellation highlight the superiority of the proposed DoTD algorithm over baseline schemes. The algorithm outperforms the benchmarks across all metrics of interest and in every scenario evaluated.

\textit{\underline{Future work:}} This study applies OSPF routing over the created network topologies. One line of future work is exploring other emerging hierarchical routing algorithms~\cite{hierarchical-routing}, or to optimize OSPF weights~\cite{ospfweights}. Another line of work is to optimizing end-to-end routes whilst incorporating elements such as ground station locations~\cite{abubakar2024choosing} and middleboxes such as CDNs~\cite{bose2024s}.

\section*{Acknowledgments}
This work is partially supported by the National Science Foundation (NSF) under Award \#2235140 and the NGIAtlantic.eu project within the EU Horizon 2020 programme under Grant No. 871582.

\end{document}